\documentclass[aps,pra,twocolumn]{revtex4}
\usepackage{graphicx}
\usepackage{bm}
\usepackage{subfig}
\usepackage{amsmath,amssymb,amsthm,dsfont,bm}
\usepackage{color}
\usepackage{wasysym}
\usepackage{ulem}
\usepackage{verbatim}
\usepackage{soul}

\usepackage[applemac]{inputenc}
\usepackage[dvipsnames]{xcolor}
\usepackage[colorlinks=true,urlcolor=blue,citecolor=blue,linkcolor=blue]{hyperref}

\begin{document}

\title{Distance-based approach to quantum coherence and nonclassicality}
\author{Laura Ares and Alfredo Luis}
\affiliation{Departamento de \'Optica, Facultad de Ciencias
F\'{\i}sicas, Universidad Complutense, 28040 Madrid, Spain}
\date{\today}

\begin{abstract}
We provide a coherence-based approach to nonclassical behavior by means of distance measures. We develop a quantitative relation between coherence and nonclassicality quantifiers, which establish the nonclassicality as the maximum quantum-coherence achievable. We compute the coherence of several representative examples and discuss whether the theory may be extended to reference observables with continuous spectra. 
\end{abstract}

\maketitle

\section{Introduction}

Coherence is the concept behind the wave nature of light and the quantum nature of physics. In quantum mechanics this is well illustrated by the Schr\"{o}dinger cats as the coherent superposition of macroscopically incompatible situations. When the coherence of the superposition vanish all quantum features disappear replaced by just classical-like ignorance of the cat state. Actually, decoherence is the most popular mechanism to account for the emergence of the classical world \cite{ZU}.  

\bigskip
This is a research area of fast grow in quantum and classical optics. In classical optics the interest has been motivated in recent times by the extension of interference-related phenomena to vector light \cite{CC1,CC2,CC3,CC4,CC5}. In quantum optics this research has been prompted by the revelation of coherence as a footing for emerging quantum technologies such as quantum information processing \cite{Li17}, and quantifying coherence has become a central task as expressed by resource theories \cite{BP14,SP17}. 

\bigskip
From this understanding of coherence as the distinctive quantum feature, it seems reasonable to assume it as the basis of any approach to nonclassical behavior from first principles. In this work we develop a quantitative relation between quantum coherence and nonclassicality. We find nonclassicality as the maximum coherence that a field state can display by varying the basis, in the same understanding that the degree of polarization is the maximum coherence between two filed modes that can be reached under unitary transformations \cite{LU08,LU10b,JA19}. 

\bigskip
The quantifier of coherence based on the {\it l1-norm} has been established as a good measure of coherence in spaces of finite dimension \cite{BP14,SP17}. In this work we express this coherence measure in terms of a Hellinger-like distance. We also define the quantifiers of all the magnitudes involved by means of this distance. In Sec. II we establish these quantifiers and derive the relation between them for finite dimensional spaces. In Sec. III we compute the coherence of some relevant states. In Sec. IV, the analysis is reproduced in infinite dimensional spaces. In Sec. V we investigate whether the theory may be extended to reference observables with continuous spectra. Finally, in Sec. VI  it is shown how these results can be replicated by using the Hilbert-Schmidt distance.

\color{black}

\bigskip

\section{Coherence quantified by a Hellinger-like  distance}

We begin the analysis with the case of an abstract space of finite dimension $N$. It is worth noting that we focus on a basis-dependent approach to coherence, so we fix a given orthogonal basis $\{|j\rangle\}_{j=1,\ldots, N}$ representing some physical variable or observable $J$, as presented for example in Ref. \cite{HE05}. 
The quantifiers utilized in this section are based on a suitable version of the Hellinger distance between two density matrices $a$ and $b$ \cite{FI16}, which is
\begin{equation}
\label{Hld}
    d_H(a,b)=\sqrt{\mathrm{tr}[(\sqrt{a}-\sqrt{b})^2]} ,
\end{equation}
where throughout this work the meaning of the square root is
\begin{equation}
\label{sqr}
    \langle i|\sqrt{a}|j \rangle=\sqrt{\langle i|a|j\rangle} ,
\end{equation}
which is slightly different from the usual definition of square root of a matrix. We can mention a similar appearance of square roots in classical-optics coherence problems \cite{MM84,MM86}.

\bigskip

Accordingly, we establish the quantifier of coherence 
based on the Hellinger distance \cite{Ji18} as the distance to the closest incoherent state $\rho_d$,
\begin{equation}
            \label{CH1}
             \mathcal{C}_{H} = \left[d_{H}(\rho,\rho_d)\right]^2 =\mathrm{tr} \left [ \left ( \sqrt{\rho}-\sqrt{\rho_d} \right)^2 \right] ,
\end{equation}
where by incoherent we mean states diagonal in the reference basis, so that $\rho_d$ turns out to be the diagonal part of $\rho$ in the same basis \cite{HE05}
\begin{equation}
\label{rhodnumber}
\rho_d = \sum_{j=1}^N \rho_{j,j} | j \rangle \langle j | ,
\end{equation} 
where $\rho_{i,j} = \langle i |\rho | j \rangle $ are the matrix elements of $\rho$ in the basis $\{|j\rangle\}$.
This quantifier can be expressed as
\begin{equation}
\label{CH2}
\mathcal{C}_H=    \displaystyle\sum_{j \neq k} | \rho_{j,k} |,
\end{equation}
which coincides with the well established quantifier of coherence $C_{l_1}$ \cite{BP14, SP17}
. This definition relies on the idea that the coherence of any state in a given basis is essentialy determided by the nondiagonal terms of its density matrix, which are obviously base-dependent. Then, if $\rho$ is diagonal in the basis $\{| j \rangle\}$ it is incoherent in such a basis $C_{H_{min}}=0$. The maximum value,  $\mathcal{C}_{H_{max}} = N-1$, can be easily computed for pure states with $\rho_{ii} = 1/N$, those are phase-like states as we discuss around Eq. (\ref{phs}). This bound is actually general beyond pure states as shown in Refs. \cite{CH15, YDXS16}.  

 An useful expression for $\mathcal{C}_H$ valid for pure states is 
\begin{equation}
\label{sCHps}
    \mathcal{C}_H = \left ( \sum_j \sqrt{p_j} \right )^2 - 1 ,
\end{equation}
where $p_j= \rho_{j,j}$ is the statistics of the basis variable $J$. 
\bigskip

In line with the distance-based measures of quantumness from quantum resource theories \cite{Ch19, Do00, Tan19} we utilize the  distance in Eqs. (\ref{Hld}) and (\ref{sqr}) to define a quantifier of nonclassicality:
\begin{equation}
\label{NCH}
    \mathcal{NC}_{H}=\left [d_{H}(\rho,I/N)\right ]^2=\mathrm{tr} \left [ \left ( \sqrt{\rho}-I/\sqrt{N} \right )^2 \right ] .
\end{equation}
As the state of reference or {\it classical state} we consider the maximally mixed state, $I/N$, since it has been shown in Ref. \cite{LM17} that under very generic conditions the normalized identity is actually the only classical state \cite{HF17}. We may invoke also the approach to nonclassicality in Ref. \cite{SP18}. These two ideas merge recalling that the identity is the only matrix which is diagonal in all bases. 
However, as well as the previously introduced measure of coherence, $\mathcal{NC}_{H}$ is also a base-dependent quantity. The minimum $\mathcal{NC}_{H_{min}} =0$ clearly holds if and only if $\rho = I/N$ as the only classical state. On the other hand, the maximum value is $\mathcal{C}_{H_{max}} = N-1$, and holds again for pure phase-like states for which $\rho_{ii} = 1/N$.

\bigskip

Finally, we quantify the fluctuations of the observable defined by the 
basis $\{|j \rangle\}$ whose probability outcomes are the diagonal terms in $\rho$, $p_j= \rho_{j,j}$. We refer to this quantity as {\it certainty} \cite{LU01,LU03} and we require it to be minimum when the probabilities are equally distributed, this is $p_j=1/N$, and maximum when the probability distribution has only one therm $p_j=\delta_{j,j_0}$.

Since these fluctuations are absolutely independent on the coherence terms of $\rho$, we define the certainty quantifier as the distance between $\rho_d$ and $I/N$: 

\begin{equation}
\label{SH}
\mathcal{S}_H(\rho)=\left [d_{H}(\rho_d,I/N)\right ]^2= \mathrm{tr} \left [ \left ( \sqrt{\rho_d}-I/\sqrt{N} \right )^2 \right ],
\end{equation}
with
\begin{equation}
\label{oSH}
     \mathrm{tr} \left [ \left ( \sqrt{\rho_d}-I/\sqrt{N} \right )^2 \right ]=2 \left ( 1-\frac{1}{\sqrt{N}}  \displaystyle\sum_{j=1}^N \sqrt{\rho_{jj}} \right ).
\end{equation}

This means that $I/N$ is the incoherent state with larger indetermination on $J$ fluctuations in the coherence basis $\{|j\rangle\}$ and therefore the larger the distance between $\rho_d$ and $I/N$ the lesser the fluctuations. As it was required, the maximum $\mathcal{S}_H (\rho) = 2 (1-1/\sqrt{N})$ holds for the elements of the coherence basis $\{|j\rangle\}$ while the minimum $\mathcal{S}_{H_{min}} = 0$ occurs when $\rho_{ii}=1/N$. Moreover,after Eq. (\ref{oSH})  we may relate $\mathcal{S}_H$ to the R\`enyi entropy of order $1/2$ \cite{Re61} 
\begin{equation}
H_{1/2}= 2 \ln \left(   \displaystyle\sum_{k=1}^n p_{k}^{1/2} \right ) .
\end{equation}

\subsection{Pythagorean equation}

{\it Theorem.} Given the previous definitions (\ref{CH1}), (\ref{NCH}) and (\ref{SH}), it can be established the following relation between magnitudes
\begin{equation}
\label{R2FW}
\textbf{Non classicality} = \textbf{Coherence} + \textbf{Certainty},
\end{equation}{}
this is 
\begin{equation}
\mathcal{NC}_{H} =  \mathcal{C}_{H} +  \mathcal{S}_{H} .
\end{equation}{}

{\it Proof.} We insert the closest incoherent state $\rho_d$ in the definition of non-classicality's quantifier in Eq. (\ref{NCH}) as $\mathrm{tr} \left [ \left ( \sqrt{\rho} - \sqrt{\rho_d} + \sqrt{\rho_d}- I/\sqrt{N} \right )^2 \right ]
$ so that it equals to
\begin{eqnarray}
   &\mathcal{NC}_{H} = \mathrm{tr} \left [ \left ( \sqrt{\rho}-\sqrt{\rho_d} \right )^2 \right ] + \mathrm{tr} \left [ \left ( \sqrt{\rho_d}-I/\sqrt{N} \right )^2 \right ] \nonumber  \\
   &+ 2 \mathrm{tr} \left [ \left ( \sqrt{\rho}-\sqrt{\rho_d} \right ) \left ( \sqrt{\rho_d}-I/\sqrt{N}\right ) \right ] .
\end{eqnarray}

As long as $\mathrm{tr} \left ( \sqrt{\rho} \sqrt{\rho_d} \right ) =  \mathrm{tr} \left (  \sqrt{\rho_d}^2 \right )$, it can be readily shown that 
\begin{equation}
\label{ort}
\mathrm{tr} \left [ \left ( \sqrt{\rho}-\sqrt{\rho_d} \right ) \left ( \sqrt{\rho_d}-I/\sqrt{N}\right ) \right ] = 0.
\end{equation}
Therefore, we obtain the following Pythagoras-like equation in a finite dimensional space:
\begin{eqnarray}
\label{R2F}
&\mathrm{tr} \left [ \left ( \sqrt{\rho}- I/\sqrt{N} \right )^2 \right ] =\\
&\mathrm{tr} \left [ \left ( \sqrt{\rho}-\sqrt{\rho_d} \right )^2 \right ] + \mathrm{tr} \left [ \left ( \sqrt{\rho_d}- I/\sqrt{N}\right )^2 \right ]. \blacksquare\nonumber
\end{eqnarray}{}

The central point of this derivation is the interpretation that we can make of each term in the underlying right-triangle structure associated to this version of the Phytagorean theorem. The hypotenuse represents nonclassicality , while coherence and certainty are the cathetus, that are orthogonal as shown in (\ref{ort}). This may be illustrated with the aid of Fig. 1, where $\rho_d$ is the orthogonal projection of $\rho$ into the incoherence hyperplane. Note that we may obtain arbitrary Pythagoras theorems replacing $I/N$ by any incoherent state, so that the orthogonality (\ref{ort}) will still hold. But the choice $I/N$ is clearly the one where hypotenuse and cathetus have a most clear physical meaning.  Note that (\ref{R2F}) and (\ref{R2FW}) adopts also the form of a duality relation between coherence and certainty in the coherence basis, as already studied in Refs. \cite{SD01,LU08} . It also worth noting that this quantum result parallels equivalent results in classical optics \cite{LU10b}.

\begin{figure}[h]
    \centering
    \includegraphics[width=8cm]{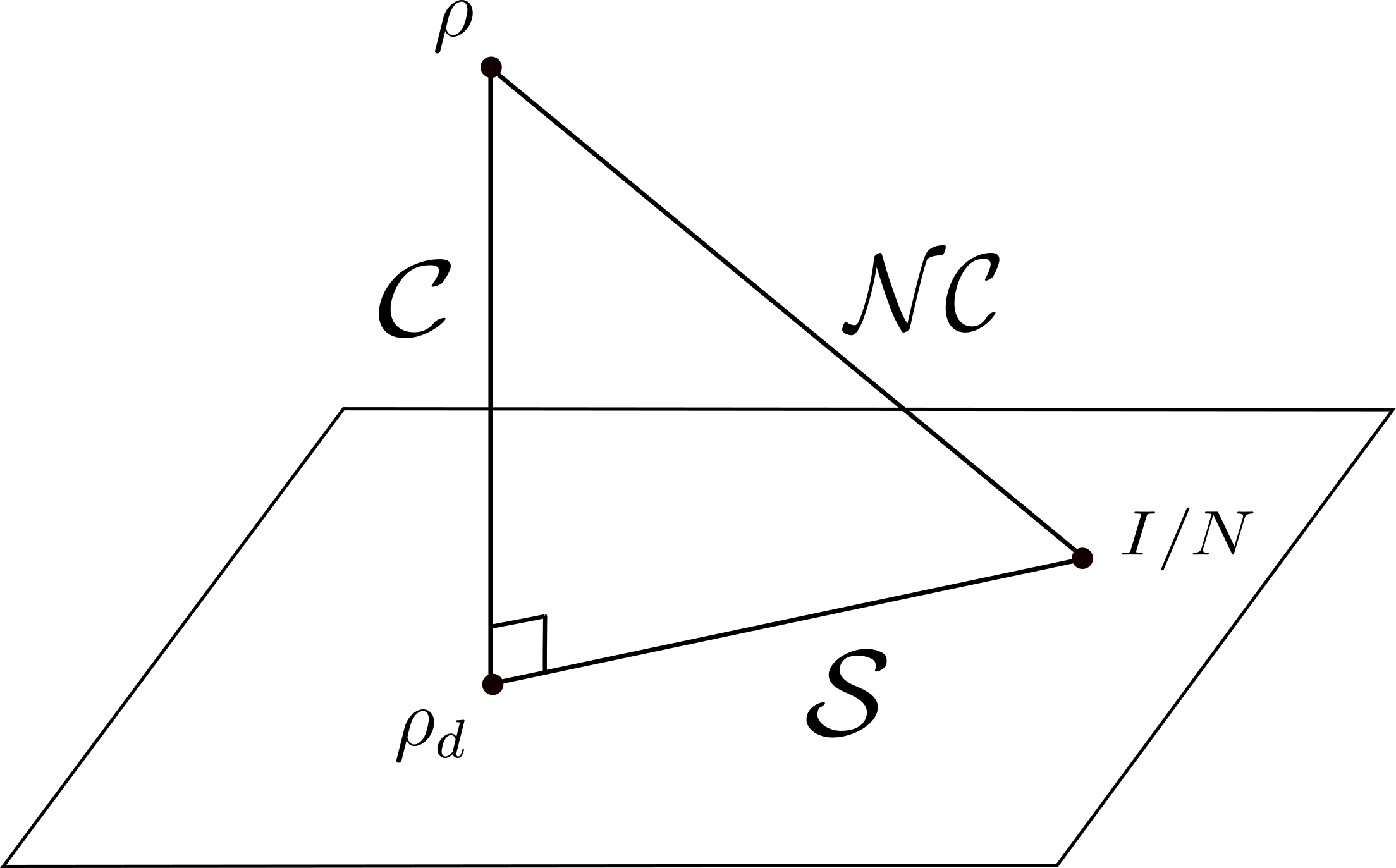}
    \caption{Pythagoras-like theorem in finite dimension.}
    \label{CHSN1f}
\end{figure}

As a direct conclusion from the previous theorem we can see that nonclassicality $\mathcal{NC}_{H}$ becomes the maximum value achievable for the coherence $\mathcal{C}_{H}$, in agreement with \cite{ST18}. In addition, the difference between them is attributed to the properties of the basis at hand. This is an interesting combination, since from a more classical-like perspective $\mathcal{NC}_{H}$ has be seen as an intrinsic or absolute form of coherence, this is independent of any reference observable \cite{Ra19,JA19,Fan16,Xu16}, see also Ref. \cite{Fe10}. 

\bigskip

By simplifying (\ref{R2F}) we arrive to the following relation between the coherence and the equivalent {\it purity} of the square root density matrix
\begin{equation}
\label{pure}
\mathrm{tr}
\left ( \sqrt{\rho}^2 \right )   
= \mathcal{C}_H + 1.
\end{equation}

Considering this simplification we try to find a different relation between coherence and certainty which does not involve the nonclassicality term. If we denote 
\begin{equation}
    x = \sum_{i=1}^N \sqrt{\rho_{ii}},
\end{equation}
then
\begin{equation}
  \mathcal{C}_H= \sum_{i\neq j}|\rho_{i j}| \leq x^2-1 , \quad {S}_H = 2 \left ( 1-\frac{x}{\sqrt{N}}  \right ) ,
\end{equation}
and the equality in $\mathcal{C}_H$ holds for pure states under the form  (\ref{sCHps}). These quantities are combined to obtain a new relation between them, arriving to:
\begin{equation}
 1 = \frac{\mathcal{S}_H}{2} + \frac{x}{\sqrt{N}}  \geq \frac{\mathcal{S}_H}{2} + \sqrt{\frac{C_H+1}{N}}
\end{equation}
where the equality holds for pure states.

\section{Examples}
Next we compute the $\mathcal{C}_H$ coherence of some meaningful states within the area of quantum optics. 

\subsection{Qubit}

This is the case $N=2$. In quantum optics the most famous qubit is a single photon split into two field modes, representing typically two orthogonal polarization states, or the two inner paths in a two-beam interferometer. A qubit can be fully characterized by three-dimensional real Bloch vector $\bm{s}$ with $|\bm{s}| \leq 1$, equivalent to the Stokes parameters if we are within a polarization context, such that 
\begin{equation}
    \rho= \frac{1}{2} \left ( 1 + \bm{s} \cdot \mathbf{\sigma} \right ),
\end{equation}
where $\mathbf{\sigma}$ are the Pauli matrices. Choosing the basis $J$ as the eigenvectors of the $\sigma_z$ matrix we have 
\begin{equation}
\mathcal{C}_H = \sqrt{s_x^2+s_y^2}, \quad \mathcal{S}_H = 2 - \sqrt{1+s_z}-\sqrt{1-s_z} ,
\end{equation}
and naturally $\mathcal{NC}_H=\mathcal{C}_H+\mathcal{S}_H$.

\bigskip

In order to look for maximum coherence and nonclassicality varying the basis, we equivalently vary the Bloch vector without altering its modulus. It can be easily seen that the maximum of both $\mathcal{C}_H$ and $\mathcal{NC}_H$ holds when the projection of the Bloch vector along the direction of the basis, say $J=\sigma_z$, vanishes, this is $s_z=0$, giving the following maxima
\begin{equation}
    \mathcal{NC}_H = \mathcal{C}_H = |\bm{s}|.
\end{equation}

\subsection{Phase states}

As already noticed, the states that make $\mathcal{C}_H$ maximum should be pure states with the same value of $\rho_{ii} = 1/N$ for all $i$, leading to $S_H =0$ and $\mathcal{NC}_{H} = \mathcal{C}_{H} = N-1$. The corresponding states are 
\begin{equation}
    \label{phs}
    |\psi \rangle = \frac{1}{\sqrt{N}} \sum_{j=1}^N e^{i\phi_j} |j \rangle ,
\end{equation}
where $\phi_j$ are arbitrary phases. With a proper phase adjustment we may say that these are finite-dimensional phase states \cite{FDPS,FDPS2} .

\subsection{Rotated number states}

Beam splitting is a traditional form of creating coherence after incoherent states in classical and quantum optics, being the basis of interferometry. So let us examine the coherence gained when incoherent number states $|n\rangle |m \rangle$ illuminate a lossless beam splitter. In particular we focus on the optimum case of a 50 \% beam splitters in the sense of providing maximum coherence. Since we consider energy-conserving processes and a finite number of photons $NT=n+m$, for all practical purposes the system is described by finite-dimensional spaces of dimension $N= NT +1$, being isomorphic to an spin $s=N/2$. These states includes SU(2) coherent states as the case $m=0$ \cite{CS}, and the Holland-Burnett states of maximum SU(2) squeezing and maximum interferometric resolution as the twin photon states $|n\rangle |n \rangle$ \cite{HB}. 

\bigskip

The 50 \% beam splitter induces a suitable mode transformation from the input modes $a,b$ to the output modes $a_1,a_2$, some phases irrelevant for our purposes, 
\begin{equation}
\label{mt}
    a_1 = \frac{1}{\sqrt{2}} \left ( a+ b \right ), \quad  a_2 = \frac{1}{\sqrt{2}} \left ( a - b \right ),
\end{equation}
such that the input state in modes $a,b$
\begin{equation}
    |n\rangle |m \rangle = \frac{1}{\sqrt{n!m!}} a^{\dagger n}b^{\dagger m}|0,0 \rangle, 
\end{equation}
transforms into the following state in the output modes $a_1,a_2$
\begin{equation}
    |n\rangle |m \rangle = \frac{1}{\sqrt{2^{m+n}n!m!}} \left ( a_1^\dagger + a_2^\dagger \right )^n \left ( a_1^\dagger - a_2^\dagger \right )^m|0,0 \rangle, 
\end{equation}
leading to 
\begin{equation}
    |n\rangle |m \rangle = \sum_{j=0}^{n+m} c_j |j \rangle ,
\end{equation}
where $|j \rangle$ are photon-number states on the modes  $a_1,a_2$, $|j \rangle = |j \rangle_1 |n+m-j \rangle_2$, and omitting an irrelevant phase, 
\begin{equation}
    c_j = \frac{\sqrt{n!m!}}{\sqrt{2^{m+n}}}\sum_{k=0}^j\frac{(-1)^k \sqrt{j!(n+m-j)!}}{k!(j-k)!(n-k)!(m+k-j)!} .
\end{equation}
 We start comparing the coherence of the SU(2) coherent states $m=0$ and the twin states $n=m$, by using Eq. (\ref{sCHps}) with $p_j= |c_j|^2$.
\begin{figure}[h]
    \centering
    \includegraphics[width=8cm]{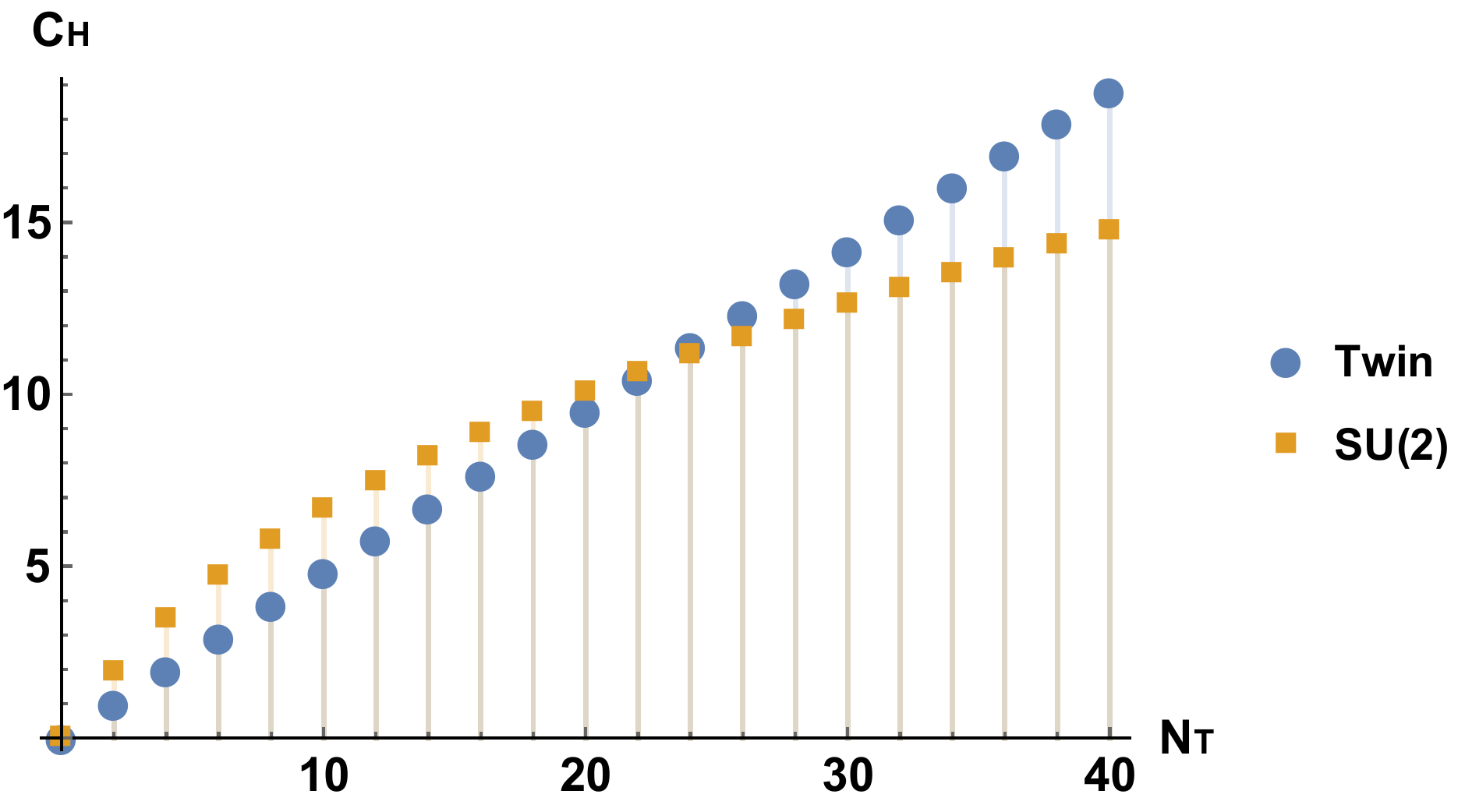}
    \caption{Coherence based on Hellinger distance of SU(2) coherent  states  and  Twin  states  as  a  function  of  the  total number of photons}
    \label{HTvsC}
\end{figure}

It can be seen in Fig. \ref{HTvsC} how the total amount of coherence increases in both cases with the total number of photons. The SU(2) coherent states are also more coherent when total number of photons is low and less coherent than the twin states when the energy of the states increases.

\bigskip

\begin{figure}[h]
    
    \centering
    \includegraphics[width=8cm]{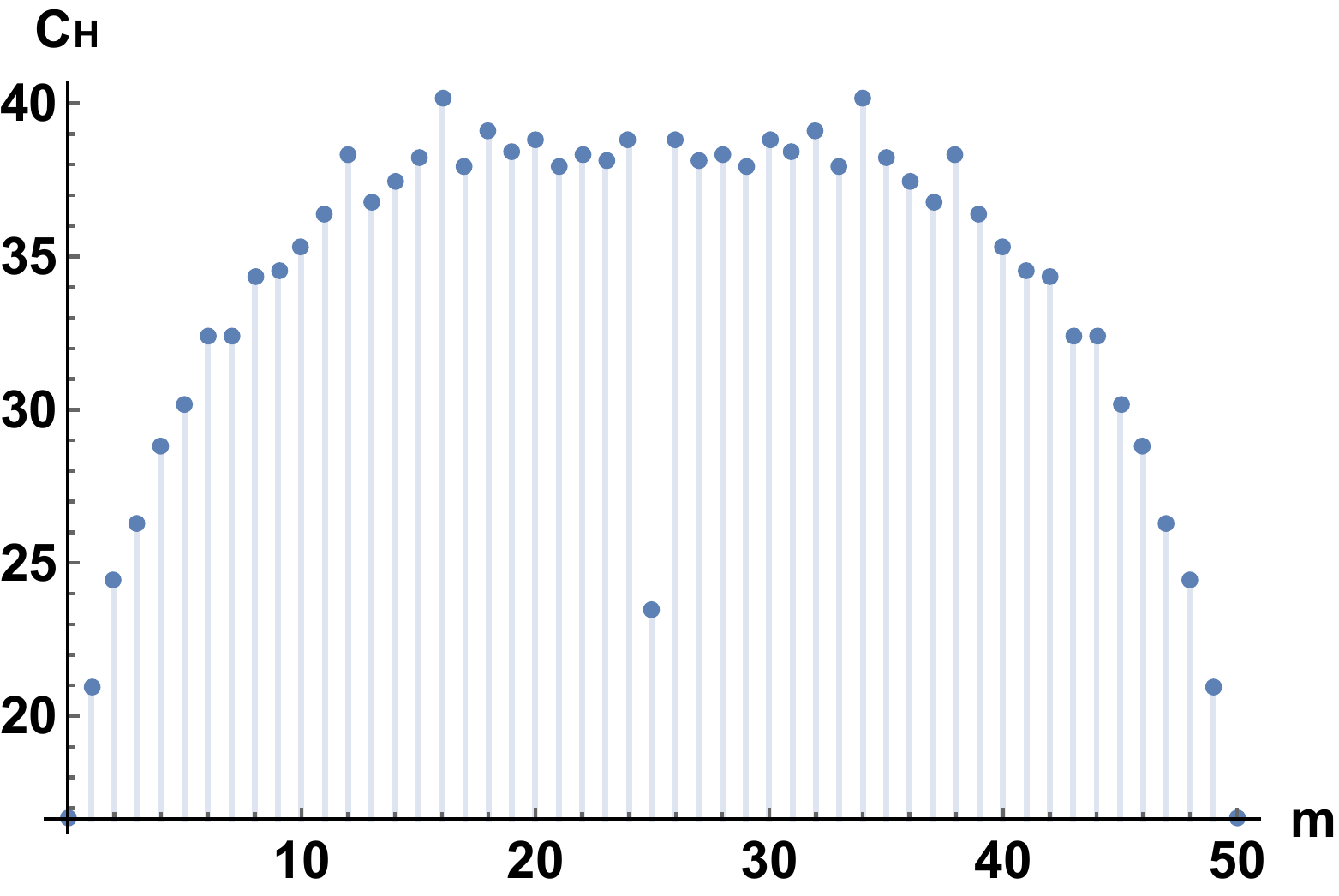}
    \caption{Coherence of rotated number states $|n \rangle |m \rangle$ as a function of $m$ for total number of photons $NT=n+m=50$ showing a general trend in which coherence increases when $n \simeq m \simeq NT/2$, except the exact equality.  }
    \label{RotatedH}
\end{figure}

For fixed $n+m$ there is a general trend in which coherence tends to be maximum around equal splitting of the photons between the input modes $n \simeq m \simeq NT/2$, curiously except the exact equality $n=m$ that shows a clear coherence dip, as displayed in Fig. \ref{RotatedH}. 

\bigskip

\section{Infinite dimension: numerable basis}
Now we extend the previous analysis to a Hilbert space of infinite dimension. We choose a numerable basis, $\{|n \rangle\}_{n=0,1, \ldots \infty}$, representing for example number of photons. The case of observables with continuous bases is examined separately below.
The translation to this area of the finite-dimensional analysis made above finds a major difficulty. This is that there can be no physical state proportional to the identity. This is to say that in infinite dimension there are no classical states. 

As discussed after Eq. (\ref{R2F}), we may expect that the Pythagorean theorem will still hold replacing the identity by any incoherent state, but the point is the physical interpretation of the terms. Because of this, in this context we replace the identity by a incoherent physical state $\rho_T$ as close as desired to have a uniform distribution in the coherence basis $\{| n \rangle\}$, this is approaching to be a maximally mixed state. This can be the case of a thermal-like state in the limit when the analog of the temperature tends to infinity

\begin{equation}
\label{HCID}
\rho_T = (1 - \xi )  \displaystyle\sum_{n=0}^\infty \xi^n | n\rangle \langle n |\quad \text{with } \xi \rightarrow 1.
\end{equation}{}

\subsection{Pythagorean equation}

So let us derive the infinite-dimensional version of the \ Pythagorean theorem (\ref{R2FW}) illustrated in Fig. 1.
As the key point of the derivation we have the orthogonality condition, 
\begin{equation}
\label{ortCont}
    \mathrm{tr} \left [ \left ( \sqrt \rho-\sqrt{\rho_d} \right ) \left ( \sqrt{\rho_d} -\sqrt{\rho_T} \right ) \right ] = 0,
\end{equation}

for the same meaning of $\rho_d$ as the diagonal part of $\rho$ in the number basis. The above relation holds for any $\rho_T$ diagonal in the number basis since 
\begin{eqnarray}
    & \mathrm{tr} \left ( \sqrt\rho \sqrt\rho_d \right ) =  \mathrm{tr} \left (  \sqrt\rho_d\sqrt{\rho_d} \right ), & \nonumber \\
     & & \\
& \mathrm{tr} \left ( \sqrt\rho \sqrt{\rho_T} \right ) = \mathrm{tr} \left ( \sqrt{\rho_T} \sqrt{\rho_d} \right ).& \nonumber
\end{eqnarray}
Therefore, we readily get this new version of Pythagoras theorem in a infinite-dimension Hilbert space:
\begin{eqnarray}
\label{PytHSI}
    &\mathrm{tr} \left [ \left ( \sqrt \rho-\sqrt{\rho_T} \right )^2 \right ] = &\\
    &\mathrm{tr} \left [ \left ( \sqrt \rho-\sqrt{\rho_d} \right )^2 \right ] + \mathrm{tr} \left [ \left ( \sqrt{\rho_d}-\sqrt{\rho_T} \right )^2 \right ], & \nonumber
\end{eqnarray}
which has the same interpretation as in the finite-dimension scenario:
\begin{equation}
\textbf{Non classicality} = \textbf{Coherence} + \textbf{Certainty}.
\end{equation}
as far as we consider the above mentioned limit for $\rho_T$ in Eq. (\ref{HCID}). Let us compute the certainty and simplify this expression as follows, 
\begin{equation}
    \mathcal{S}_H =\mathrm{tr}  \left ( \sqrt {\rho_d}^2 \right )  +\mathrm{tr} \left ( \sqrt {\rho_T}^2 \right ) -2\mathrm{tr}\left ( \sqrt {\rho_d} \sqrt \rho_T  \right ) ,
\end{equation}
  with
\begin{align}
&\mathrm{tr} \left ( \sqrt{\rho_d}^2 \right ) = \mathrm{tr} \left ( \sqrt{\rho_T}^2 \right ) = 1 ,&
\end{align}

while for the third term we have
\begin{align}
&\mathrm{tr} \left ( \sqrt{\rho_d} \sqrt{\rho_T} \right ) = \sqrt{1 - \xi}   \displaystyle\sum_{n=0}^\infty \xi^{n/2} \sqrt{p_n}, &
\end{align}
 where $p_n = \langle n | \rho| n \rangle$. In order to proceed with the $\xi \rightarrow 1$ limit we shall consider that the following quantity is finite, which is   the key ingredient of coherence as shown in Eq. (\ref{sCHps}), 
\begin{equation}
\label{cond}
    \sum_{n=0}^\infty \sqrt{p_n} < \infty .
\end{equation}
This is actually satisfied by all the cases to be considered in this work. In such a case, when  $\xi \rightarrow 1$ we get 
\begin{equation}
\mathrm{tr} \left ( \sqrt{\rho_d} \sqrt{\rho_T} \right ) \rightarrow 0, 
\end{equation}
so that $\mathcal{S}_H=2$ and 
\begin{equation}
\label{nr}
\mathcal{NC}_H  
= C_H + 2.
\end{equation}
Roughly speaking, $\mathcal{S}_H=2$ means that as $\xi \rightarrow 1$ the distance between the {\it physical} state $\rho_d$ and $\rho_T$ tends to be maximum.
Therefore, it is worth noting that in this infinite dimensional case we get that, in the conditions specified above, coherence equals nonclassicality.

\subsection{Examples}

\subsubsection{Number states}
As the elements of the coherence basis they are incoherent having maximum certainty
\begin{equation}
    \mathcal{C}_H=0, \quad \mathcal{S}_H =\mathcal{NC}_H = 2 .
\end{equation}

\subsubsection{Phase states}

In the case of the normalizable Susskind-Glogower phase states \cite{SGPS}
\begin{equation}
    | \xi \rangle = \sqrt{1-|\xi|^2 }\sum_{n=0}^\infty \xi^n | n \rangle,
\end{equation}
the coherence becomes
\begin{equation}
\label{phaseCH}
    \mathcal{C}_H= \frac{1 +|\xi|}{1-|\xi|}-1 = \frac{2|\xi|}{1-|\xi|} ,
\end{equation}
 so that $\mathcal{C}_H \rightarrow \infty$ as $|\xi | \rightarrow 1$. In terms of the mean number of photons 
\begin{equation}
    \bar{n} = \frac{|\xi |^2}{1 - |\xi |^2} , \quad  |\xi |^2 = \frac{\bar{n}}{1+\bar{n}} ,  
\end{equation}
we have 
\begin{equation}
\label{CHTMPS}
    \mathcal{C}_H 
    = 2 \left ( \bar{n} + \sqrt{ \bar{n}+\bar{n}^2}\right ),
\end{equation}
that for large enough $\bar{n} \gg 1$ the coherence scales linearly with the mean number of photon as 
\begin{equation}
    \mathcal{C}_H \simeq 4 \bar{n}.
\end{equation}

\subsubsection{Two-mode squeezed vacuum}

The results for the phase sates can be easily translated to the case of the two-mode squeezed vacuum because of their form similarity

\begin{equation}
\label{tmsv}
    |\xi \rangle = \sqrt{1-|\xi|^2} \sum_{n=0}^\infty \xi^n |n, n\rangle ,
\end{equation}
made just of twin-photon states we will obtain the same expression for the coherence as in (\ref{phaseCH}), which in this case means the more squeezing the more coherence.

\bigskip

\subsubsection{ Squeezed coherent states.}
We compute the coherence in the photon-number basis of pure displaced squeezed vacuum states,  with displacement or coherent amplitude $R$, and squeezing parameter $r$. 

\bigskip

\begin{figure}[h]
    \centering
    \subfloat
   {\includegraphics[width=8cm]{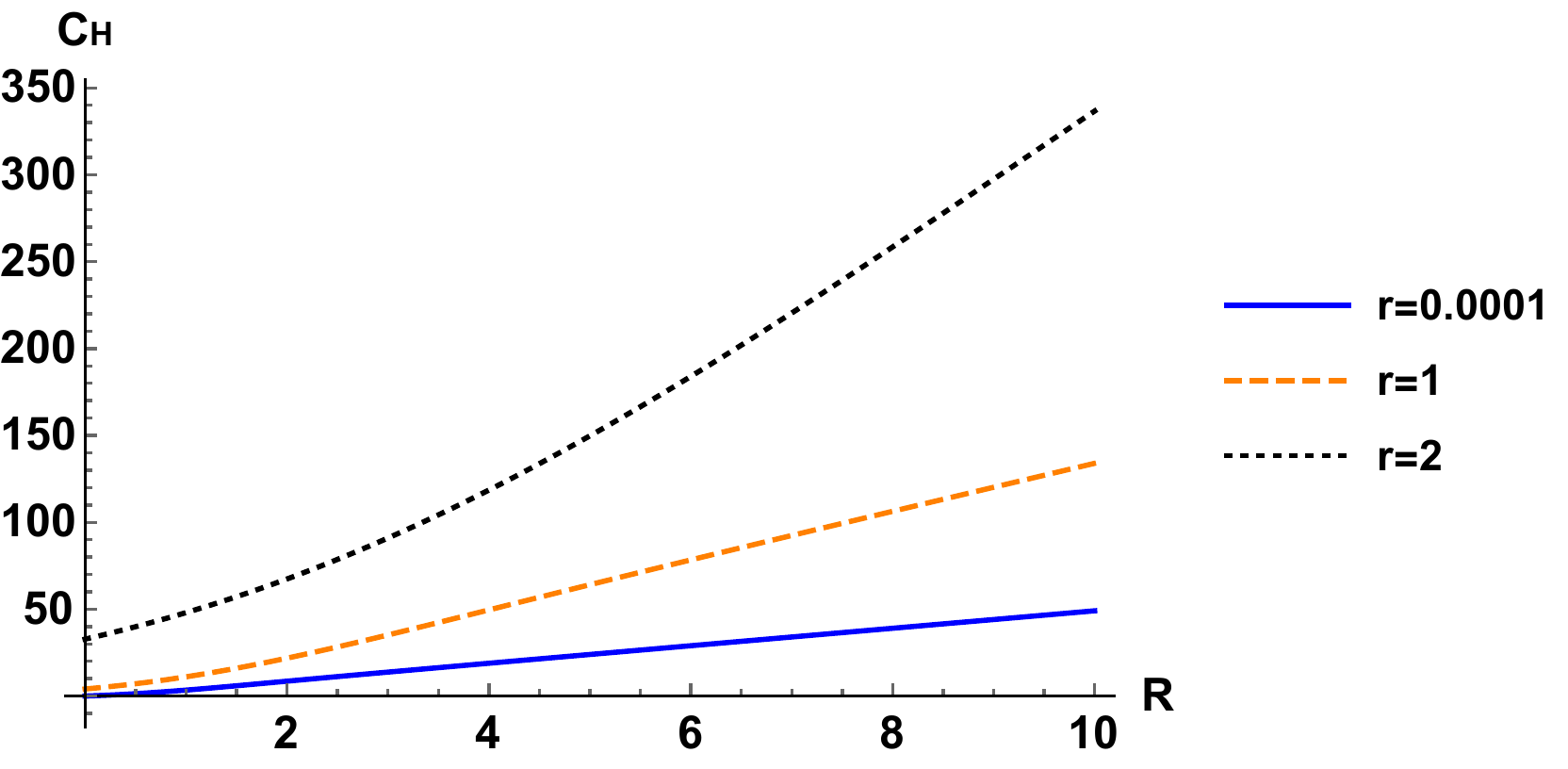}}
    \caption{Coherence of squeezed coherent states with different compression parameters, as a function of the displacement, $R$. }
    \label{CHDn-R}
    \subfloat
   {\includegraphics[width=8cm]{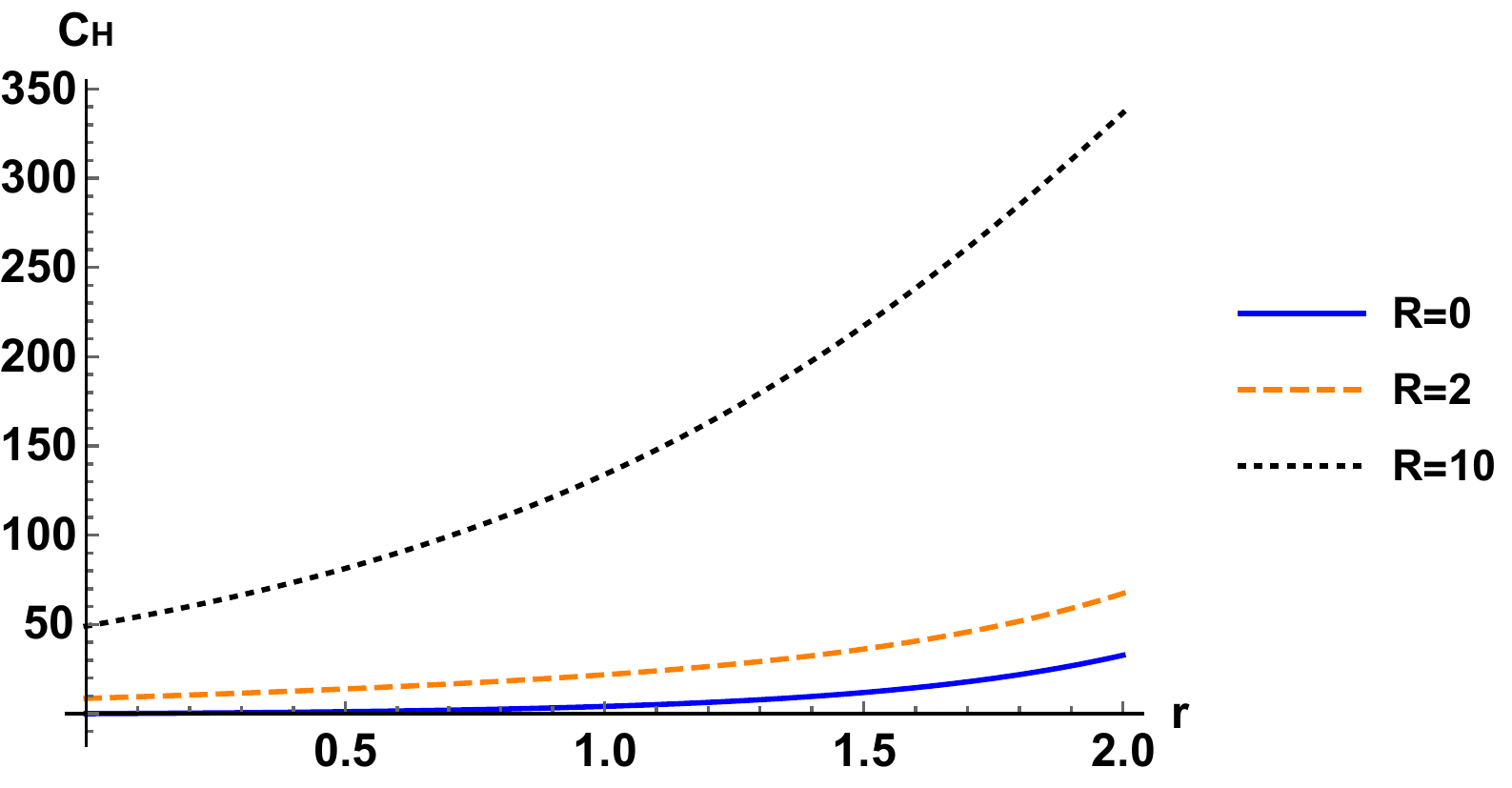}}
   \caption{Coherence of squeezed coherent states with different coherent displacements as a function of the squeeze parameter, $r$.}
    \label{CHDn-r2}
\end{figure}

In Fig. \ref{CHDn-R} it is shown how the larger the displacement $R$, the larger the coherence, almost in a linear way. In Fig. \ref{CHDn-r2} it can be seen how coherence raises with the compression parameter, $r$, then squeezed coherent states have more coherence than coherent states.

This behaviour can be understood recalling that for large displacements $R$ and not too large squeezing the photon-number distribution  of squeezed coherent states can be well approximated by a continuous Gaussian distribution. In such a case, after the suitable generalization of Eq.(\ref{sCHps}) to this situation, the coherence can be readily computed to give 
\begin{equation}
\label{Ga}
    \mathcal{C}_H \simeq 2 \sqrt{2 \pi \Delta^2 n}-1 ,
\end{equation}
where $\Delta n$ is the number uncertainty, which in these conditions might be approximated on the form $\Delta^2 n \simeq \bar{n} e^{2r}$. So we see that coherence increases with the photon-number variance, and that squeezing can always increase fluctuations via super-Poissonian number statistics. 

\bigskip

\begin{figure}
\includegraphics[width=8cm]{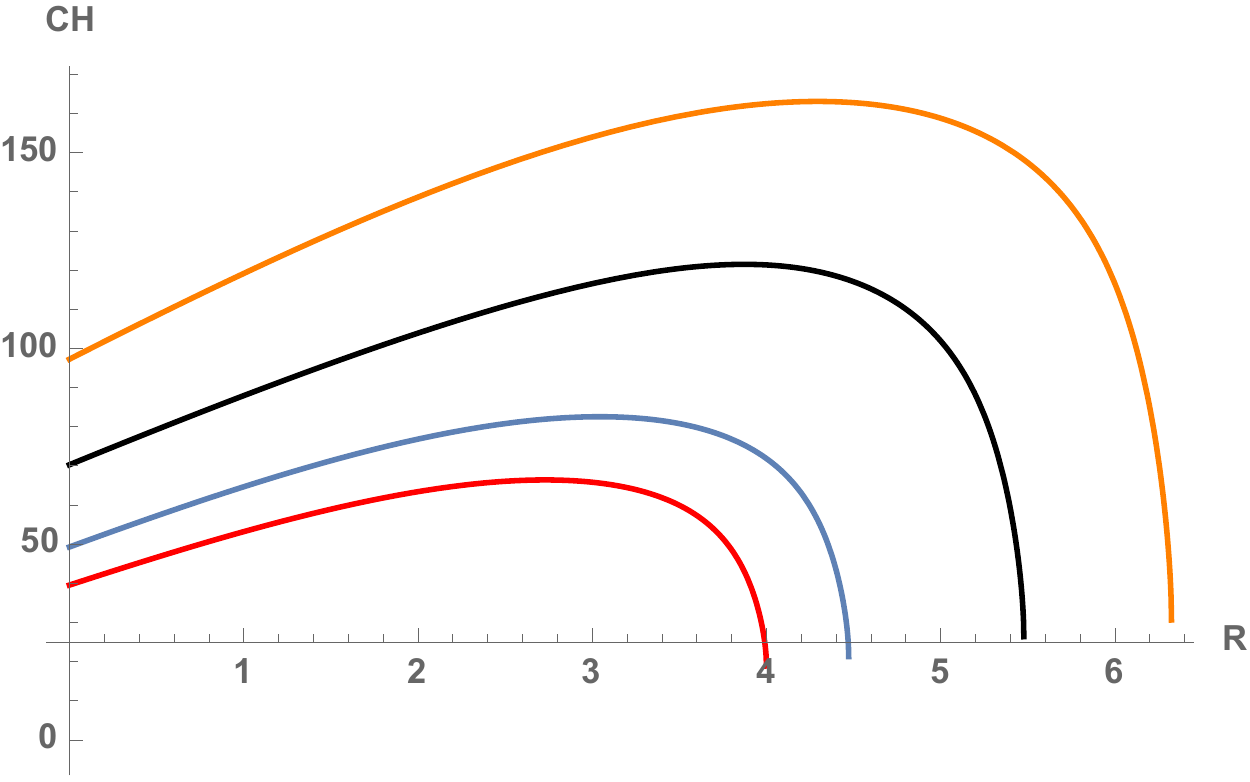}
\caption{Coherence of squeezed coherent states as a function of the displacement $R$ for determined mean number of photons, $\bar{n}=16$ red line, $\bar{n}=20$ blue line, $\bar{n}=30$ black line, $\bar{n}=40$ orange line. }
    \label{CHN30}
\end{figure}

Finally, we fix the mean photon number in order to study the optimum distribution of energy between squeezing and displacement. In Fig. \ref{CHN30} we observe an optimum distribution of this energy when around 30\% is utilized to squeeze the state. In this case, the optimum configuration supposes an important improvement in the total amount of coherence. Roughly speaking, such optimum configuration agrees with the limit in which squeezed coherent states become suitable approximations of normalized phase phase states, as states that tend to be optimum regarding metrology \cite{LU06}.

\bigskip

{}

\bigskip

\subsubsection{Displaced-number states}

\begin{figure}[h]
    \centering
    \includegraphics[width=8cm]{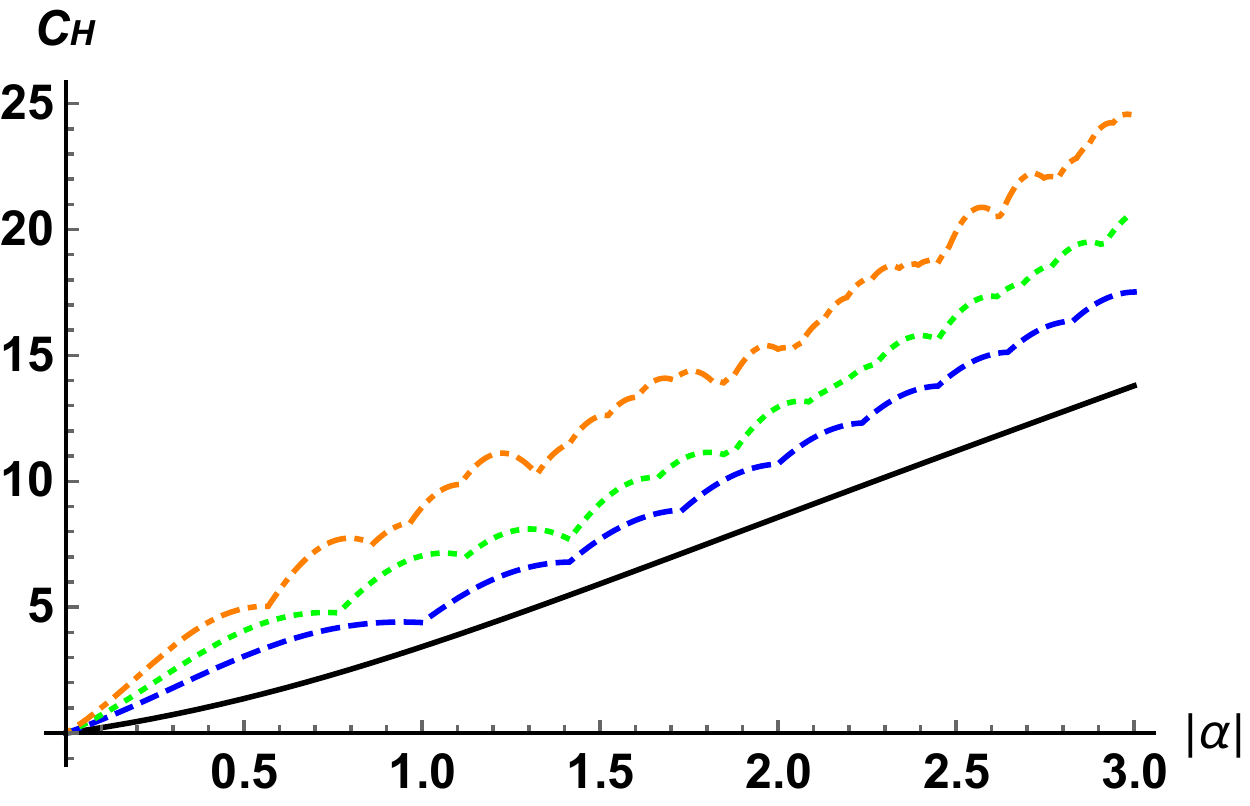}
    \caption{Coherence for displaced number states $D(\alpha ) |n_0 \rangle$ as a function of the displacement $|\alpha|$ for $n_0=0$ in solid black line, this is a coherent state, $n_0=1$ in dashed blue line, $n_0=2$ in dotted green line, and $n_0=4$ in dash-dotted orange line. }
    \label{DnH}
\end{figure}

We examine the contribution of the displacement to the coherence in the case of displace number states, $D(\alpha ) |n_0 \rangle$, where $D(\alpha ) |n_0 \rangle$ where $D(\alpha ) = \exp( \alpha a^\dagger - \alpha^\ast a )$ is the displacement operator. We find that the general trend of coherence is to grow with $|\alpha|$ as in the squeezed coherent state case. More specifically, in Fig. \ref{DnH} it is shown how this growth is softer for states the coherent state, $n_0=0$.

\section{Continuous bases}

In this section we attempt to extend the previous analysis to continuous bases both in finite and infinite-dimensional spaces.

By a suitable generalization of the preceding analyses we may consider as coherence with respect to any basis $|\phi \rangle$, even if it is continuous or nonorthogonal, the contribution of the nondiagonal terms of $\rho$, this is an expression of the form
\begin{equation}
\label{cb}
\mathcal{C}_{H}= \mathrm{tr} \left ( \sqrt{\rho}^2 \right ) -1 .
\end{equation}
taken from Eq. (\ref{pure}). The question to be addressed next is whether such definition of coherence has the same geometrical meaning we have found above in the case of discrete orthogonal basis. To discuss this we focus on whether there is a proper definition of $\rho_d$ as the incoherent state closest to $\rho$. First we consider normalized nonorthogonal bases in finite-dimensional spaces, and then orthogonal nonnormalized ones in infinite-dimensional spaces.

\color{black}

\subsection{Finite-dimensional space}
For definiteness we use as the basis the set of finite-dimensional phase states
\begin{equation}
| \phi \rangle = \frac{1}{\sqrt{N}}\sum_{j=1}^N e^{i j \phi} |j \rangle ,
\end{equation}
where the $|j \rangle$ refers to some orthonormal number-like basis.
In this scenario we may consider 
\begin{equation}
\label{rhodphase}
\rho_d = \frac{N}{2 \pi} \int d\phi \langle \phi |  \rho | \phi \rangle | \phi \rangle \langle \phi | 
\end{equation}
to be the ``incoherent" state of reference, no longer diagonal as we will see in the following. In addition, it is necessary to determine the meaning of the square root suitable for this continuous framework. Thus we define $\sqrt{\rho_d}$ as
\begin{equation}
    \sqrt{\rho_d} =\sqrt{ \frac{N}{2 \pi}} \int d\phi \sqrt{\langle \phi |  \rho | \phi \rangle }| \phi \rangle \langle \phi |,
\end{equation}
so that $C=0$ if $\rho=\rho_d$.

After these definitions it turns out that  $\mathrm{tr} \left [(\sqrt{\rho}-\sqrt{\rho_d})^2\right]$ does not reproduces Eq. (\ref{cb}) nor the Pythagorean theorem holds due to 
\begin{equation}
\mathrm{tr} \left [ \left ( \sqrt{\rho} - \sqrt{\rho_d} \right) \left ( \sqrt{\rho_d} - I/\sqrt{N} \right )\right ] \neq 0,
\end{equation}
as it can be easily checked for example for the qubit state. We may ascribe this behavior to the lack of orthogonality of the phase states  
\begin{equation}
\label{nondiagonal}
    \langle \phi^\prime | \phi \rangle = \frac{1}{N}\sum_j^N e^{i j(\phi-\phi')}\neq 0,
\end{equation}
which makes $\rho_d$ non-diagonal, meaning 
\begin{equation}
\label{nonortogonal}
\langle \phi^\prime |\rho_d |\phi \rangle \neq 0,  \quad \phi \neq \phi^\prime.
\end{equation}

\subsection{Infinite-dimensional space.}

Let us consider next the case of a continuous basis made of unnormalizable orthogonal states, such as the quadrature or position eigenstates $|x\rangle$, where $x$ can take any real value. Although they are orthogonal in the sense of $\langle x^\prime | x \rangle = \delta (x - x^\prime)$ there is the difficulty of $|x\rangle$ being no normalizable. As a consequence, any state diagonal in the $|x\rangle$ basis is not physical since its trace diverges, in particular this is the case of the following definition of $\rho_d$:
\begin{equation}
\label{rdx}
\rho_d =  \int_{-\infty}^\infty dx  \langle x |\rho | x \rangle |x\rangle \langle x |.
\end{equation} 
As we have done above with $\rho_T$ we can try to avoid this via some kind of regularization in some proper limit. 
To this end we replace $|x\rangle$ by some normalizable states, for example displaced-squeezed states $|x\rangle_\Delta$ with quadrature-coordinate wave function
\begin{equation}
\label{wf}
    \langle x^\prime|x\rangle_\Delta = \frac{1}{(2 \pi \Delta^2)^{1/4}} \exp \left [-\frac{(x-x^\prime)^2}{4 \Delta^2} \right ]
\end{equation}
so that we can define a truly unit-trace $\rho_d$ as
\begin{equation}
\label{rdxc}
\rho_d =  \int_{-\infty}^\infty dx  \langle x |\rho | x \rangle |x\rangle_\Delta {}_\Delta \langle x |
\end{equation} 
in the spirit of considering afterwards the limit $\Delta \rightarrow 0$. With this definition it can be easily seen that 
\begin{equation}
\lim_{\Delta \rightarrow 0} \langle x | \rho_d | x \rangle =  \langle x |\rho | x \rangle ,
\end{equation} 
simply by invoking the Gaussian representation of the Dirac delta function
\begin{equation}
\label{delta}
    \lim_{\Delta \rightarrow 0} |\langle x^\prime|x\rangle_\Delta |^2 = \delta (x - x^\prime ) . 
\end{equation}

Now we try to recover an expression for the coherence in Eq. (\ref{cb}) as a suitable trace-distance, this is in terms of
$\mathrm{tr}\left[ \left (\sqrt{\rho}- \sqrt{\rho_d} \right )^2 \right]$. To this end we introduce $\sqrt{\rho}$ and $\sqrt{\rho_d}$ as
\begin{equation}
\sqrt{\rho} =  \int_{-\infty}^\infty dx^\prime \int_{-\infty}^\infty dx
\sqrt{  \langle x |\rho | x^\prime \rangle} |x\rangle\langle x^\prime |
\end{equation} 
and
\begin{equation}
\label{rdsxc}
\sqrt{\rho_d} =  \int_{-\infty}^\infty dx\sqrt{  \langle x |\rho | x \rangle} |x\rangle_\Delta {}_\Delta \langle x |.
\end{equation} 
respectively.

It can be seen that $ \mathrm{tr} \left ( \sqrt{\rho_d}^2 \right )$ vanish in the limit $\Delta \rightarrow 0$ since
\begin{equation}
\label{delta2}
    \lim_{\Delta \rightarrow 0}| {}_\Delta \langle x^\prime|x\rangle_\Delta |^2 = 2\sqrt{\pi} \Delta \delta (x - x^\prime )\rightarrow 0.
\end{equation}

Similarly, taking into account Eq. (\ref{wf}), in the limit $\Delta \rightarrow 0$ we may consider that $\langle x^\prime | x \rangle_\Delta $ and ${}_\Delta \langle x | x^{\prime \prime} \rangle$ are so peaked functions so that they act as Dirac delta functions  
\begin{equation}
\label{delta3}
   \lim_{\Delta \rightarrow 0}\langle x^\prime | x \rangle_\Delta {}_\Delta \langle x | x^{\prime \prime} \rangle = 2 \sqrt{2\pi} \Delta \delta(x-x^\prime) \delta (x - x^{\prime \prime}),
\end{equation}
and therefore, in this limit

\begin{equation}
    \mathrm{tr} \left ( \sqrt{\rho_d }\sqrt{\rho} \right ) \rightarrow 2 \sqrt{2\pi} \Delta \int dx  \langle x |\rho | x \rangle \rightarrow 0.
\end{equation}

All this together it emerges that
\begin{equation}
\lim_{\Delta \rightarrow 0} \mathrm{tr} \left [ \left ( \sqrt{\rho} -\sqrt{\rho_d} \right )^2 \right ] =  \mathrm{tr}  \left ( \sqrt{\rho}^2 \right )  
\end{equation}
so Eq. (\ref{cb}) is essentially recovered. In view of this we wonder whether Pythagorean relation in Eq. (\ref{PytHSI}) also holds.
Note that in this continuous case $\sqrt{\rho_T}$ is defined as

\begin{equation}
    \sqrt{\rho_T}=\int dx \sqrt{\langle x |\rho_T| x \rangle }|x\rangle_{\Delta}{}_\Delta \langle x|.
\end{equation}
\bigskip

We answer this question in the afirmative since the limits (\ref{delta2}) and (\ref{delta3}) ensure the orthogonality condition in Eq. (\ref{ortCont}).

\section{Coherence quantified by the Hilbert-Schmidt distance}

The previous results can be reproduced by using the Hilbert-Schmidt distance to quantify all the magnitudes involved \cite{Do00,Oz00},
\begin{equation}
    d_{HS}(a,b)=\sqrt{\mathrm{tr}[(a-b)^2]}.
\end{equation}

We consider this scenario since the coherence based on the Hilbert-Schmidt distance is widely utilized \cite{BP14,SP17},
\begin{equation}
            \label{CHS}
             \mathcal{C}_{HS} (\rho) = \left [d_{HS}(\rho,\rho_d)\right ]^2=\mathrm{tr} \left [ \left ( \rho-\rho_d \right)^2 \right]=    \displaystyle\sum_{j \neq k} | \rho_{j,k} | ^2. 
\end{equation}
with $\rho_d$ defined in Eq. (\ref{rhodnumber}). This distance allows us to recover an equivalent Pythagoras-like equation in a finite dimensional space: 
\begin{equation}
\label{R2I}
\mathrm{tr} \left [ \left ( \rho- I/N \right )^2 \right ] = \mathrm{tr} \left [ \left ( \rho-\rho_d \right )^2 \right ] + \mathrm{tr} \left [ \left ( \rho_d- I/N\right )^2 \right ],\nonumber 
\end{equation}{}
\begin{equation}
    \mathcal{NC}_{HS}=\mathcal{C}_{HS}+\mathcal{S}_{HS}.
\end{equation}

The states making these quantities extremal are the same as for Hellinger quantifiers. The only difference is the maximum value of $\mathcal{NC}_{HS}$ and $\mathcal{C}_{HS}$ which becomes $1-1/N$ so in this case the coherence, nonclassicality and certainty are bounded by $1$ in finite dimensional spaces.

\bigskip

Furthermore, in infinite dimensional spaces with numerable basis we also arrive at an equivalent Pythagoras-like equation by means of same classical reference $\rho_T$ introduced in Eq. (\ref{HCID}),

\begin{equation}
    \mathrm{tr} \left [ \left ( \rho-\rho_T \right )^2 \right ] = \mathrm{tr} \left [ \left ( \rho-\rho_d \right )^2 \right ] + \mathrm{tr} \left [ \left ( \rho_d-\rho_T \right )^2 \right ]. 
\end{equation}{}

We complete the analysis of the Hilbert-Schmidt scenario with the translation into continuous basis.
\bigskip

In the case of finite dimensional spaces and nonortogonal basis, the difficulties caused by the definition of the closest incoherent state (see Eqs. (\ref{nondiagonal}) and (\ref{nonortogonal})) remain, so it is also impossible to find a suitable geometrical formulation of coherence and nonclassicality in such continuous framework.

Finally we consider an infinite dimensional space and continuous bases made of unnormalizable orthogonal  states, where the very same definition of the incoherent state proposed in Eq.(\ref{rdxc}) can be utilized. As a result of the limits in Eqs. (\ref{delta2}) and (\ref{delta3}) we arrive at
\begin{equation}
 \lim_{\Delta \rightarrow 0} \mathrm{tr} \left [ \left ( \rho -\rho_d \right )^2 \right ]=   \mathrm{tr}  \left ( \rho^2 \right ) ,
\end{equation}
which is a base-independent quantity. Thus, in this case of Hilbert-Schmidt distance the translation into continuous basis is not possible neither for infinite nor for finite dimensional spaces.

\color{Black}

\bigskip

\section{Conclusions}
We have carried out a study of coherence based on distance measurements. We have developed a relation between coherence, certainty and nonclassicality which establishes the latter as the upper bound of the coherence, ascribing the difference to the basis at hand. This relation can be extended to infinite dimensional systems through discrete basis as well as continuous orthogonal basis. We conclude that there is no straightforward expansion of the formalism to the case of continuous non-orthogonal bases. All these conclusions are shared by the analyses made with both Heillinger-like and Hilbert-Schmidt distances, except from the case of continuous non-orthogonal basis, since there is no proper expansion to continuous basis in any case when using the Hilbert-Schmidt distance.

The examples examined show the increase of coherence with squeezing on squeezed coherent states and with the displacement on coherent and number states. Also we find an interesting and abrupt reduction in the coherence of twin states.

\bigskip
\bigskip
\bigskip

\noindent{\bf Acknowledgments.- }
L. A. and A. L. acknowledge financial support from Spanish Ministerio de Economia y Competitividad Project No. FIS2016-75199-P.
L. A. acknowledges financial support from European Social Fund and the Spanish Ministerio de Ciencia Innovacion y Universidades, Contract Grant No. BES-2017-081942.

\end{document}